\documentclass[twocolumn,prd,showpacs,preprintnumbers,amsmath,amssymb,floatfix]{revtex4}

\usepackage{graphicx}

\usepackage{bm}
\usepackage{amsfonts}
\usepackage{lineno,hyperref}
\usepackage{array}
\usepackage{microtype}
\usepackage{float}

\newcommand*\DAlambert{\mathop{}\!\mathbin\Box}

\begin{document}

\title{Gravastars in $f(R,\mathcal{T})$ gravity}

\author{Amit Das}
\email{amdphy@gmail.com}
\affiliation{Department of Physics, Indian Institute of Engineering
Science and Technology, Shibpur, Howrah 711103, West Bengal,
India}

\author{Shounak Ghosh}
\email{shnkghosh122@gmail.com}
\affiliation{Department of Physics, Indian Institute of Engineering
Science and Technology, Shibpur, Howrah 711103, West Bengal,
India}

\author{Swapan Das}
\email{swapan.d11@gmail.com}
\affiliation{Department of Optometry, NSHM College of Management and Technology,
124 B.L. Saha Road, Kolkata 700053, West Bengal, India}

\author{Farook  Rahaman}
\email{rahaman@associates.iucaa.in}
\affiliation{Department of Mathematics, Jadavpur University, Kolkata
700032, West Bengal, India}

\author{B.K. Guha}
\email{bkguhaphys@gmail.com}
\affiliation{Department of Physics, Indian
Institute of Engineering Science and Technology, Shibpur, Howrah
711103, West Bengal, India}

\author{Saibal Ray}
\email{saibal@associates.iucaa.in}
\affiliation{Department of Physics, Government College of Engineering and Ceramic Technology, Kolkata 700010, West Bengal, India}

\date{\today}

\begin{abstract}
We propose a unique stellar model under the $f(R,\mathcal{T})$
gravity by using the conjecture of Mazur-Mottola~[P. Mazur and E.
Mottola, Report number: LA-UR-01-5067., P. Mazur and E. Mottola,
Proc. Natl. Acad. Sci. USA {\bf 101}, 9545 (2004).] which is known
as gravastar and a viable alternative to the black hole as
available in literature. This gravastar is described by the three
different regions, viz., (I) Interior core region, (II)
Intermediate thin shell, and (III) Exterior spherical region. The
pressure within the interior region is equal to the constant
negative matter density which provides a repulsive force over the
thin spherical shell. This thin shell is assumed to be formed by a
fluid of ultrarelativistic plasma and the pressure, which is
directly proportional to the matter-energy density according to
Zel'dovich's conjecture of stiff fluid~[Y.B. Zel'dovich, Mon. Not.
R. Astron. Soc. {\bf 160}, 1 (1972).], does counterbalance the
repulsive force exerted by the interior core region. The exterior
spherical region is completely vacuum and assumed to be de Sitter
spacetime which can be described by the Schwarzschild solution.
Under this specification we find out a set of exact and
singularity-free solution of the gravastar which presents several
other physically valid features within the framework of
alternative gravity.
\end{abstract}


\maketitle

\section{Introduction}
Mazur and Mottola~\cite{Mazur2001,Mazur2004} first ever proposed a
model considering the {\it gra}vitationally {\it va}cuum {\it
star} (gravastar) as an alternative to the system of gravitational
collapse, i.e., black hole. They generated a new type of solution
by extending the idea of Bose-Einstein condensation in
construction of the gravastar as a cold, dark, and compact object
of interior de Sitter condensate phase.

The scenario of this gravastar can be envisaged as follows: the
interior is surrounded by a thin shell of ultrarelativistic matter
whereas the exterior region is completely vacuum and hence the
Schwarzschild spacetime at the outside can be considered to fit
for the system. The shell is assumed to be very thin with a finite
width in the range $ r_1 < r < r_2 $, where $r_1 \equiv D$ and
$r_2 \equiv D+\epsilon$ are the interior and exterior radii of the
gravastar respectively under consideration. Therefore, we can
represent the entire system of gravastar into three specific
segments based on the equation of state (EOS) as follows: (I)
Interior ($0 \leq r < r_1 $):~$p = -\rho $, (II) Shell ($ r_1 \leq
r \leq r_2 $):~$ p = +\rho $, and (III) Exterior ($ r_2 < r $):~$
p = \rho =0$.

We note that related to the gravastar there are lot of works
available in the literature based on different mathematical as
well as physical issues. However, these works are mainly carried
out by several authors in the framework of Einstein's general
relativity~\cite{Mazur2001,Mazur2004,Visser2004,Cattoen2005,Carter2005,Bilic2006,Lobo2006,DeBenedictis2006,Lobo2007,Horvat2007,Cecilia2007,Rocha2008,Horvat2008,Nandi2009,Turimov2009,Usmani2011,Lobo2013,Bhar2014,Rahaman2015}.
Though it is well known that Einstein's general relativity is a
unique tool for uncovering many hidden mysteries of Nature, yet
some observational evidences of the accelerating universe along
with the existence of dark matter has imposed a theoretical
challenge to this theory
~\cite{Ri1998,Perl1999,Bern2000,Hanany2000,Peebles2003,Paddy2003,clifton2012}.
Therefore, several alternative theories have been proposed
successively amongst which $f(R)$ gravity, $f(\mathbb{T})$
gravity, and $f(R,\mathcal{T})$ gravity have received more
attention. In the present project our motivation is to study the
gravastar under one of the alternative gravity theories, namely
$f(R,\mathcal{T})$ gravity~\cite{harko2011} and to observe
different physical features of the object - their nontriviality as
well as triviality. Actually our previously performed successful
works on the initial phases of compact stars under alternative
gravity~\cite{Das2015,Das2016} motivate us to exploit the
alternative formalism to the case of the gravastar, a viable
alternative to the ultimate stellar phase of a black hole.

It has been argued that among all other modified gravity theories
the $f(R,\mathcal{T})$ theory of gravity can be considered as a
useful formulation which is based on the nonminimally curvature
matter coupling. In the $f(R,\mathcal{T})$ theory of
gravity~\cite{harko2011} the gravitational Lagrangian of the
standard Einstein-Hilbert action is defined by an arbitrary
function of the Ricci scalar $R$ and the trace of the
energy-momentum tensor $\mathcal{T}$. One can note that such a
dependence on $\mathcal{T}$ may come from the presence of an
imperfect fluid or from the consideration of quantum effects. The
application of $f(R,\mathcal{T})$ gravity theory to different
cosmological~\cite{moraes2014b,moraes2015a,moraes2015b,singh2014,rudra2015,baffou2015,shabani2013,shabani2014,sharif2014b,reddy2013b,kumar2015,shamir2015,Fayaz2016}
realm can be noted in the literature.

Among several astrophysical applications it is worthy of
mentioning the
Refs.~\cite{sharif2014,noureen2015,noureen2015b,noureen2015c,zubair2015a,zubair2015b,Ahmed2016,Moraes2016,Yousaf2016a,Yousaf2016b}.
In their work~\cite{sharif2014} Sharif et al. have studied the
stability of collapsing spherical body of an isotropic fluid
distribution considering the nonstatic spherically symmetric line
element. A perturbation scheme has been used to find the collapse
equation and the condition on the adiabatic index has been
constructed for Newtonian and post-Newtonian eras for addressing
instability problem by Noureen et al.~\cite{noureen2015} whereas
in another work ~\cite{noureen2015b} Noureen et al. have
investigated the range of instability under the $f(R,\mathcal{T})$
theory for an anisotropic background constrained by zero
expansion. Also, by applying a perturbation scheme on the
$f(R,\mathcal{T})$ field equations the evolution of a spherical
star has been studied by Noureen et al. ~\cite{noureen2015c}.
Zubair et al.~\cite{zubair2015a} have analyzed the dynamics of
gravitating sources along with axial symmetry under the
$f(R,\mathcal{T})$ gravity. Some other relevant studies on the
$f(R,\mathcal{T})$ theory of gravity can be observed in the
following works~\cite{zubair2015b,Ahmed2016,Moraes2016} under
different physical motivations. Yousaf et al.~\cite{Yousaf2016a}
have explored the evolutionary behaviors of compact objects in the
framework of $f(R,\mathcal{T})$ gravity theory with the help of
structure scalars whereas they~\cite{Yousaf2016b} have
investigated the irregularity factors for a self-gravitating
spherical star evolving in the presence of imperfect fluid.

The outline of the present study is therefore as follows: In
Sec.~II the basic mathematical formalism of the $f(R,\mathcal{T})$
theory has been provided as the background of the study.
Thereafter in Sec.~III we discuss the field equations and their
solutions in $f(R,\mathcal{T})$ gravity considering the interior
spacetime, exterior spacetime, and thin shell cases of the
gravastars with their respective solutions. We provide the
junction conditions, which are essential in connection to the
three regions of the gravastar, in Sec.~IV. Several physical
properties of the models, viz. proper length, energy content,
entropy and equation of state, are discussed in Sec.~V. Some
concluding remarks are provided in Sec.~VI.

\section{Basic mathematical formalism of the $f(R,\mathcal{T})$ Theory}
The action of the $f(R,\mathcal{T})$ theory~\cite{harko2011} reads
\begin{equation}\label{eq1}
\mathbb{S}=\frac{1}{16\pi}\int
d^{4}xf(R,\mathcal{T})\sqrt{-g}+\int d^{4}x\mathcal{L}_m\sqrt{-g},
\end{equation}
where $f(R,\mathcal{T})$ is the function of the Ricci scalar $R$
and the trace of the energy-momentum tensor $\mathcal{T}$,
$\mathcal{L}_m$ being the matter Lagrangian density, and $g$ is
the determinant of the metric $g_{\mu\nu}$. Throughout the paper
we assume the geometrical units $G=c=1$.

Varying the action (\ref{eq1}) with respect to the metric
$g_{\mu\nu}$, one can obtain the following field equations of
$f(R,\mathcal{T})$ gravity:
\begin{eqnarray}\label{eq2}
f_R (R,\mathcal{T}) R_{\mu\nu} - \frac{1}{2} f(R,\mathcal{T}) g_{\mu\nu}
+ (g_{\mu\nu}\DAlambert - \nabla_{\mu} \nabla_{\nu}) f_R (R,\mathcal{T})\nonumber \\
= 8\pi T_{\mu\nu} - f_\mathcal{T}(R,\mathcal{T}) T_{\mu\nu} -
f_\mathcal{T}(R,\mathcal{T})\Theta_{\mu\nu},
\end{eqnarray}
where $f_R (R,\mathcal{T})= \partial f(R,\mathcal{T})/\partial R$,
$f_\mathcal{T}(R,\mathcal{T})=\partial f(R,\mathcal{T})/\partial
\mathcal{T}$, $\DAlambert \equiv
\partial_{\mu}(\sqrt{-g} g^{\mu\nu} \partial_{\nu})/\sqrt{-g}$,
$R_{\mu\nu}$ is the Ricci tensor, $\nabla_\mu$ provides the
covariant derivative with respect to the symmetric connection
associated to $g_{\mu\nu}$, $\Theta_{\mu\nu}=
g^{\alpha\beta}\delta T_{\alpha\beta}/\delta g^{\mu\nu}$ and the
stress-energy tensor is defined as
$T_{\mu\nu}=g_{\mu\nu}\mathcal{L}_m-2\partial\mathcal{L}_m/\partial
g^{\mu\nu}$.

The covariant divergence of (\ref{eq2}) reads
~\cite{barrientos2014}
\begin{eqnarray}\label{eq3}
\hspace{-0.5cm}\nabla^{\mu}T_{\mu\nu}&=&\frac{f_\mathcal{T}(R,\mathcal{T})}{8\pi -f_\mathcal{T}(R,\mathcal{T})}[(T_{\mu\nu}+\Theta_{\mu\nu})\nabla^{\mu}\ln f_\mathcal{T}(R,\mathcal{T}) \nonumber \\
&&+\nabla^{\mu}\Theta_{\mu\nu}-(1/2)g_{\mu\nu}\nabla^{\mu}\mathcal{T}].
\end{eqnarray}

It is vivid from Eq. (\ref{eq3}) that the energy-momentum tensor is
not conserved for the $f(R,\mathcal{T})$ theory of gravity unlike
the general relativistic case.

In the present paper we assume the energy-momentum tensor to be
that of a perfect fluid, i.e.,
\begin{equation}\label{eq4}
T_{\mu\nu}=(\rho+p)u_\mu u_\nu-pg_{\mu\nu},
\end{equation}
with $u^{\mu}u_{\mu} = 1$ and $u^\mu\nabla_\nu u_\mu=0$. Besides these
conditions we also have $\mathcal{L}_m=-p$ and
$\Theta_{\mu\nu}=-2T_{\mu\nu}-pg_{\mu\nu}$.

Following the  proposition of Harko et al.~\cite{harko2011}, we
take the functional form of $f(R,\mathcal{T})$ as
$f(R,\mathcal{T})=R+2\chi \mathcal{T}$, with $\chi$ being a
constant. One can note that this form has been extensively used to
obtain many cosmological solutions in $f(R,\mathcal{T})$
gravity~\cite{singh2015,moraes2014b,moraes2015a,moraes2015b,reddy2013b,kumar2015,shamir2015}.
By substituting the above form of $f(R,\mathcal{T})$ in
(\ref{eq2}), we get~\cite{moraes2014b,moraes2015a}
\begin{equation}\label{eq5}
G_{\mu\nu}=8\pi T_{\mu\nu}+\chi \mathcal{T}
g_{\mu\nu}+2\chi(T_{\mu\nu}+pg_{\mu\nu}),
\end{equation}
where $G_{\mu\nu}$ is the Einstein tensor.

One can easily get back to the result of general relativity just
by setting $\chi=0$ in the above Eq. (\ref{eq5}). Moreover, for
$f(R,\mathcal{T})=R+2\chi \mathcal{T}$, Eq. (\ref{eq3}) yields
\begin{equation}\label{eq6}
\nabla^{\mu}T_{\mu\nu}=-\frac{2\chi}{(8\pi+2\chi)}\left[\nabla^{\mu}(pg_{\mu\nu})+\frac{1}{2}g_{\mu\nu}\nabla^{\mu}\mathcal{T}\right].
\end{equation}

Curiously, by substituting $\chi=0$ in Eq. (\ref{eq6}) one can verify that
the energy-momentum tensor is conserved as in the case of general
relativity.

\section{The field equations and their solutions in $f(R,\mathcal{T})$ gravity}
For the spherically symmetric metric
\begin{equation}
ds^2=e^{\nu(r)}dt^2-e^{\lambda(r)}dr^2-r^2(d\theta^2+\sin^2\theta
d\phi^2),\label{eq7}
\end{equation}
one can find the nonzero components of the Einstein tensors as
\begin{equation}
G_0^{0}=\frac{e^{-\lambda}}{r^{2}}(-1+e^{\lambda}+\lambda'
r),\label{eq8}
\end{equation}

\begin{equation}
G_1^{1}=\frac{e^{-\lambda}}{r^{2}}(-1+e^{\lambda}-\nu'
r),\label{eq9}
\end{equation}

\begin{equation}
G_2^{2}=G_3^{3}=\frac{e^{-\lambda}}{4r}[2(\lambda'-\nu')-(2\nu''+\nu'^{2}-\nu'\lambda')r],\label{eq10}
\end{equation}
where primes stand for derivative with respect to the radial
coordinate $r$.

Substituting Eqs. (4), (\ref{eq8}), (\ref{eq9}), and (\ref{eq10})
in Eq. (\ref{eq5}) one can get
\begin{equation}
-1+e^{\lambda}+\lambda'r=\Pi(r)[8\pi\rho+\chi(3\rho-p)],\label{eq11}
\end{equation}

\begin{equation}
-1+e^{\lambda}-\nu'r=\Pi(r)[-8\pi p+\chi(\rho-3p)], \label{eq12}
\end{equation}

\begin{equation}
[\frac{r}{2}(\lambda'-\nu')-(2\nu''+\nu'^{2}-\nu'\lambda')\frac{r^2}{4}]=\Pi(r)[-8\pi
p+\chi(\rho-3p)],\label{eq13}
\end{equation}
with $\Pi(r)\equiv r^{2}/e^{-\lambda}$.

Now, from the equation for the nonconservation of the
energy-momentum tensor in $f(R,\mathcal{T})$ theory (\ref{eq6})
one can obtain
\begin{equation}
\frac{dp}{dr}+\frac{\nu'}{2}(\rho+p)+\frac{\chi}{2(4\pi+\chi)}(p'-\rho')=0.\label{eq14}
\end{equation}

If we consider the quantity $m$ as the gravitational mass
within the sphere of radius $r$, then from Eq. (11) we can write
\begin{equation}
e^{-\lambda}=1-\frac{2m}{r}-\chi(\rho-\frac{p}{3})r^2.\label{eq15}
\end{equation}

Again from Eqs. (12), (14), and (15) one can get the equation
of hydrostatic equilibrium in $f(R,\mathcal{T})$ theory as
\begin{equation}
p'=-(\rho+p)\frac{\left[ 4\pi
pr-\frac{\chi(\rho-3p)r}{2}\right]+\frac{1}{2r}\left[\frac{2m}{r}
+\chi(\rho-\frac{p}{3})r^2\right]}{\left[1-\frac{2m}{r}
-\chi(\rho-\frac{p}{3})r^2\right]\left[1+\frac{\chi(1-\frac{d\rho}{dp})}
{2(4\pi+\chi)}\right]},\label{eq16}
\end{equation}
considering the fact that the energy density $\rho$ depends on the
pressure $p$ i.e. $\rho=\rho(p)$.

Also, by letting $\chi=0$ the standard form of the
Tolman-Oppenheimer-Volkoff (TOV) equation can be retrieved as
applicable in the case of general theory of relativity.

\subsection{Interior spacetime}
Following the proposition of
Mazur-Mottola~\cite{Mazur2001,Mazur2004}, let us assume the
equation of state (EOS) for the interior region as
\begin{equation}
p=-\rho.\label{eq17}
\end{equation}

The above EOS is a special form of $p=\omega\rho$,
with the EOS parameter $\omega=-1$ and is known as the
dark energy equation of state.

Again using the above EOS, and from Eq. (\ref{eq14}) one can obtain
\begin{equation}
\rho = \rho_0~(constant),\label{eq18}
\end{equation}
and the pressure turns out to be
\begin{equation}
p=-\rho_0.\label{eq19}
\end{equation}

Now, using Eqs. (\ref{eq11}) and (\ref{eq19}) one gets the metric
potential $\lambda$ as
\begin{equation}
e^{-\lambda} =
1-\frac{4(2\pi+\chi)\rho_0r^2}{3}+\frac{A}{r},\label{eq20}
\end{equation}
where $A$ is an integration constant which is set to zero as the
solution is regular at the center ($r=0$). Hence we have
\begin{equation}
e^{-\lambda} = 1-\frac{4(2\pi+\chi)\rho_0r^2}{3}. \label{eq21}
\end{equation}

Again, using Eqs. (\ref{eq11}), (\ref{eq12}),  (\ref{eq18}) and
(\ref{eq19}) one can get the following relation between the metric
potentials $\nu$ and $\lambda$ as
\begin{equation}
e^{\nu}=Be^{-\lambda}, \label{eq22}
\end{equation}
where $B$ is an integration constant. Here the spacetime metric is
free from any central singularity.

Also the gravitational mass $M(D)$ can be found out to be
\begin{equation}
M(D)= \int_0^{r_1=D} 4\pi r^2{\rho_0} dr=\frac{4}{3}\pi
D^3\rho_0.\label{eq23}
\end{equation}

\subsection{Shell}
Let us consider that the shell consists of ultrarelativistic
fluid, obeying the EOS $p=\rho$. Zel'dovich~\cite{zeldovich1972}
conceived the idea of this fluid in connection to cold baryonic
universe and is known as the stiff fluid. In the present context
this may come from thermal excitations with negligible chemical
potential or from conserved number density of gravitational quanta
at zero temperature~\cite{Mazur2001,Mazur2004}. This type of fluid
has been extensively used by several authors to study various
cosmological~\cite{carr1975,Madsen1992} as well as
astrophysical~\cite{wesson1978,braje2002,linares2004} phenomena.

One can note that within the nonvacuum region, i.e., the shell it
is very difficult to find solution of the field equations.
However, it is possible to obtain an analytical solution within
the framework of thin shell limit, i.e., $0< e^{-\lambda}\ll1$.
Physically this means that when two space-times join together at a
place (in our case the vacuum interior and the Schwarzchild
exterior) the intermediate region must be a thin shell (see the
Ref.~\cite{Israel1966}). Now in thin shell as $r\rightarrow 0$,
any parameter which is function of $r$ is, in general, $\ll1$.
Under this approximation along with the above EOS as well as Eqs.
(\ref{eq11}), (\ref{eq12}) and (\ref{eq13}), one can find the
following equations
\begin{equation}
\frac{de^{-\lambda}}{dr}=\frac{2}{r},\label{eq24}
\end{equation}

\begin{equation}
\left(\frac{3}{2r}+\frac{\nu'}{4}\right)\frac{de^{-\lambda}}{dr}=\frac{1}{r^2}.\label{eq25}
\end{equation}

Integrating Eq. (\ref{eq24}) we get
\begin{equation}
e^{-\lambda}= 2\ln r + C,\label{eq26}
\end{equation}
where $C$ is an integration constant and range of $r$ is
$D\leq\,r\leq{D+\epsilon}$. Under the condition $\epsilon\ll1$, we get $C\ll1$ as $e^{-\lambda}\ll1$.

Also from Eqs. (\ref{eq24}) and (\ref{eq25}) one can get
\begin{equation}
e^{\nu}=Fr^{-4},\label{27}
\end{equation}
where $F$ is an integration constant.

Also Eq. (\ref{eq14}), along with the EOS $p=\rho$, yields
\begin{equation}
p=\rho=Hr^{4},\label{28}
\end{equation}
$H$ being a constant. As $\rho\propto\,r^{4}$, we can infer that
the ultrarelativistic fluid within the shell is more dense at the
outer boundary than the inner boundary.

\begin{figure}[h]
\centering
\includegraphics [width=0.4\textwidth]{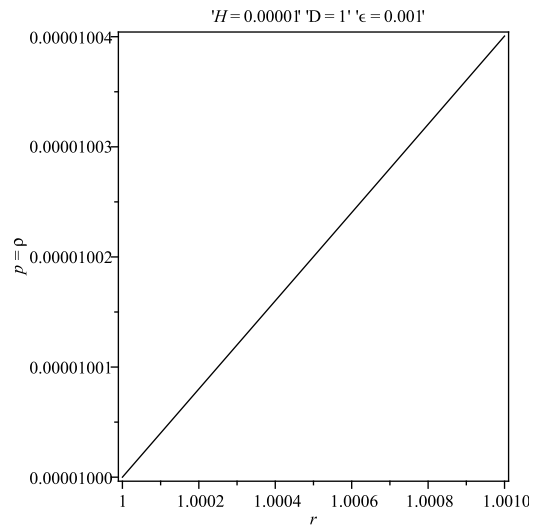}
\caption{Pressure $p=\rho$ ($km^{-2}$) of the ultrarelativistic
fluid in the shell is plotted with respect to the radial
coordinate $r$ ($km$).}
\end{figure}

\subsection{Exterior spacetime}
The exterior region obeying the EOS $(p=\rho=0)$ can be defined by the well-known static exterior
Schwarzschild solution which is given by
\begin{eqnarray}
ds^2=\left(1-\frac{2M}{r}\right)dt^2-\left(1-\frac{2M}{r}\right)^{-1}dr^2
\nonumber \\ -r^2\left(d\theta^2+sin^2\theta
d\phi^2\right),\label{eq29}
\end{eqnarray}
where $M$ is the total mass of the gravitating system.

\section{Junction Condition}
It is already mentioned that the gravastar consists of three regions, i.e., interior region (I), shell (II), and exterior region (III).
The interior region (I) is connected with the exterior region at the junction interface, i.e., at the shell.
According to the Darmois-Israel formalism~\cite{Darmois1927,Israel1966} there should be smooth
matching between the regions I and III of the gravastar. The metric coefficients are continuous
at the junction surface ($\Sigma$), i.e., at $r=D$, though their derivatives may not be continuous.
However, one can determine the surface stress-energy $S_{ij}$ by using the above mentioned formalism.

Now, the intrinsic surface stress-energy tensor $S_{ij}$ is given by the Lanczos equation~\cite{lanczos1924,sen1924,Darmois1927,Israel1966,perry1992,lake1996} as
\begin{equation}
S^i_j=-\frac{1}{8\pi}(\kappa^i_j-\delta^i_j\kappa^k_k),\label{eq30}
\end{equation}
where $\kappa_{ij}=K^+_{ij}-K^-_{ij}$ provide the discontinuity in
the second fundamental forms or extrinsic curvatures. Here the
signs ``$+$'' and ``$-$'' correspond to the interior and the
exterior regions respectively. Now, the second fundamental
forms~\cite{rahaman2006,rahaman2009,usmani2010,rahaman2010,dias2010,rahaman2011}
associated with the two sides of the shell are given by
\begin{equation}
K_{ij}^{\pm}=-n_{\nu}^{\pm}\left[\frac{\partial^{2}x_{\nu}}{\partial
\xi^{i}\partial\xi^{j}}+\Gamma_{\alpha\beta}^{\nu}\frac{\partial
x^{\alpha}}{\partial \xi^{i}}\frac{\partial x^{\beta}}{\partial
\xi^{j}} \right]|_\Sigma, \label{eq31}
\end{equation}
where $\xi^{i}$ are the intrinsic coordinates on the shell, $n_{\nu}^{\pm}$ are the unit normals to the
surface $\Sigma$ and for the spherically symmetric static metric
\begin{equation}
ds^2=f(r)dt^2-\frac{dr^2}{f(r)}-r^2(d\theta^2+sin^2\theta\,d\phi^2),\label{eq32}
\end{equation}
$n_{\nu}^{\pm}$ can be written as
\begin{equation}
n_{\nu}^{\pm}=\pm\left|g^{\alpha\beta}\frac{\partial f}{\partial
x^{\alpha}}\frac{\partial f}{\partial x^{\beta}}
\right|^{-\frac{1}{2}}\frac{\partial f}{\partial x^{\nu}},
\label{eq33}
\end{equation}
with $n^{\mu}n_{\mu}=1$.

Using the Lanczos equation we can get the surface stress-energy
tensor as {\bf $S_{ij}=diag  [{\sigma, -\upsilon, -\upsilon,
-\upsilon }$]}, where $\sigma$ is the surface energy density and
$\upsilon$ is the surface pressure. The surface energy density
($\sigma$) and the surface pressure ($\upsilon$) can be respectively expressed as
\begin{equation}
\sigma=-\frac{1}{4\pi D}\left[\sqrt{f}\right]^+_-, \label{eq34}
\end{equation}

\begin{equation}
\upsilon=-\frac{\sigma}{2}+\frac{1}{16\pi}\left[\frac{f^{'}}{\sqrt{f}}\right]^+_-.
\label{eq35}
\end{equation}

So, by using the above two equations we obtain
\begin{equation}
\sigma=-\frac{1}{4\pi D}\left[
\sqrt{1-\frac{2M}{D}}-\sqrt{1-\frac{4(2\pi+\chi)\rho_0D^2}{3}}\right],
\label{eq36}
\end{equation}

\begin{equation}
\upsilon=\frac{1}{8\pi
D}\left[\frac{(1-\frac{M}{D})}{\sqrt{1-\frac{2M}{D}}}-
\frac{\left\lbrace 1-\frac{8(2\pi+\chi)\rho_0D^2}{3}\right\rbrace
} {\sqrt{1-\frac{4(2\pi+\chi)\rho_0D^2}{3}}}\right]. \label{eq37}
\end{equation}

Also, the mass of the thin shell can be written as
\begin{equation}
m_s=4\pi D^2\sigma=D\left[
\sqrt{1-\frac{4(2\pi+\chi)\rho_0D^2}{3}}-\sqrt{1-\frac{2M}{D}}\right].
\label{38}
\end{equation}

Here $M$ is the total mass of the gravastar and it can be expressed in the following form
\begin{equation}
M=\frac{2(2\pi+\chi)\rho_0D^3}{3}+ m_s \sqrt{1-\frac{4(2\pi+\chi)\rho_0D^2}{3}}-\frac{m_s^2}{2D}.\label{39}
\end{equation}

\section{Physical features of the model}

\subsection{Proper length of the shell}
Let us consider that the stiff fluid shell is situated at the surface $r=D$ defining
the phase boundary of region I. The proper thickness of the shell is assumed to be
very small, i.e., $\epsilon\ll1$. Thus the region III starts from the interface at
$r={D+\epsilon}$. So, the proper thickness between two interfaces, i.e., of the shell is determined as
\begin{equation}
\ell= \int_D^{D+\epsilon}
\sqrt{e^{\lambda}}dr=\int_D^{D+\epsilon}\frac{dr}{\sqrt{2\ln r +
C}}.\label{eq40}
\end{equation}

Integrating the above equation one can get
\begin{equation}
\ell=\left[
-\left(-\frac{\pi}{2e^C}\right)^{\frac{1}{2}}~erf\left\lbrace\frac{\sqrt{\ln\left(\frac{1}{r^2}\right)-C}}{\sqrt{2}}\right\rbrace\right]_D^{D+\epsilon}.\label{eq41}
\end{equation}

\begin{figure}[h]
\centering
\includegraphics [width=0.4\textwidth]{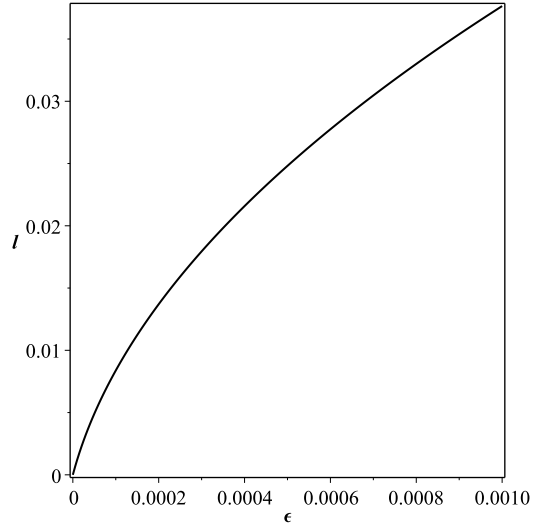}
\caption{Proper length $\ell$ ($km$) of the shell is plotted with
respect to the thickness of the shell $\epsilon$ ($km$).}
\end{figure}

\subsection{Energy content}
In the interior region we consider the EOS in the form $p=-\rho$ which indicates the negative energy
region confirming the repulsive nature of the interior region.

However, the energy within the shell turns out to be
\begin{eqnarray}
\mathcal{E}=\int_{D}^{D+\epsilon}4\pi\rho\,r^{2}dr=\int_{D}^{D+\epsilon}4\pi
H\,r^{6}dr \nonumber \\= \frac{4\pi\,H}{7}\left[{\left(
D+\epsilon\right)}^7-D^7 \right].\label{eq42}
\end{eqnarray}

Taking into account the thin shell approximation one may write the
energy $\mathcal{E}$ up to the first order of $\epsilon\,(\ll1)$
as
\begin{equation}
\mathcal{E}\approx 4\pi\epsilon\,H\,D^6.\label{eq43}
\end{equation}

\begin{figure}[h]
\centering
\includegraphics [width=0.4\textwidth]{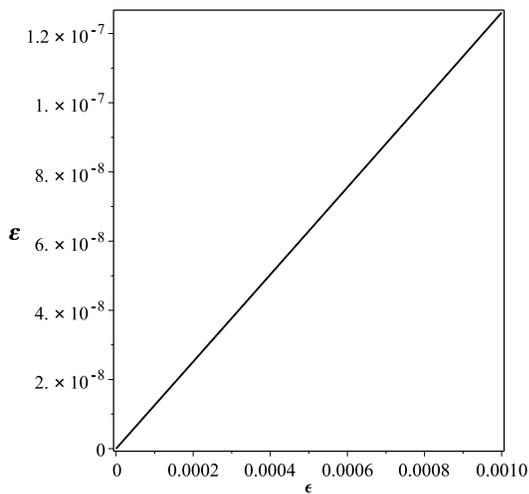}
\caption{Energy $\varepsilon$ ($km$) within the shell is plotted
with respect to the thickness of the shell $\epsilon$ ($km$).}
\end{figure}

The above relation indicates that the energy of the shell is directly proportional to the $\epsilon$, i.e., the thickness of the shell.

\subsection{Entropy}
According to the prescription of Mazur and Mottola~\cite{Mazur2001,Mazur2004} in the interior region I, the entropy density is zero which is consistent with a single condensate state. However, within the shell the entropy is given by
\begin{equation}
S=\int_{D}^{D+\epsilon}4\pi\,r^{2}s(r)\sqrt{e^{\lambda}}dr,\label{eq44}
\end{equation}
where $s(r)$ is the entropy density for local temperature $T(r)$ and may be written as~\cite{Mazur2001,Mazur2004}
\begin{equation}
s(r)=\frac{\alpha^2k_B^2T(r)}{4\pi\hbar^2 } =
\alpha\left(\frac{k_B}{\hbar}\right)\sqrt{\frac{p}{2 \pi
}},\label{eq45}
\end{equation}
$\alpha$ being a dimensionless constant. We note that in the
present work we assume the geometrical units, i.e., $G=c=1$, and
also in Planckian units $k_B=\hbar=1$. So, the entropy density
within the shell turns out to be
\begin{equation}
s(r)=\alpha\sqrt{\frac{p}{2\pi}}.\label{eq46}
\end{equation}

\begin{figure}[h]
\centering
\includegraphics [width=0.4\textwidth]{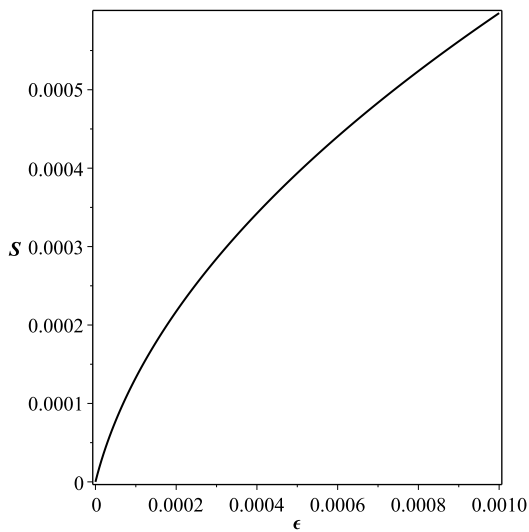}
\caption{Entropy $S$ within the shell is plotted with respect to
the thickness of the shell $\epsilon$ ($km$).}
\end{figure}

Therefore, Eq. (\ref{eq44}) can be written as
\begin{equation}
S
=\left(8\pi\,H\right)^{\frac{1}{2}}\alpha\int_{D}^{D+\epsilon}\frac{r^4}{\sqrt{2\ln
r + C}}dr.\label{47}
\end{equation}

Integrating the above equation we get
\begin{eqnarray}
S=\left(8\pi\,H\right)^{\frac{1}{2}}\alpha\left[
-\left(-\frac{\pi}{10e^{5C}}\right)^{\frac{1}{2}}\right. \times
\nonumber \\ \left.
~erf\left\lbrace\frac{\sqrt{5[\ln\left(\frac{1}{r^2}\right)
-C]}}{\sqrt{2}}\right\rbrace\right]_D^{D+\epsilon}.\label{eq48}
\end{eqnarray}

\subsection{Equation of state}
The EOS, at $r=D$, as usual can be expressed in the following form
\begin{equation}
\upsilon=\omega(D)\sigma. \label{eq49}
\end{equation}

Hence, by virtue of Eqs. (\ref{eq36}) and (\ref{eq37}) the
equation of state parameter can explicitly be written as
\begin{equation}
\omega(D)=\frac{\left[\frac{(1-\frac{M}{D})}{\sqrt{1-\frac{2M}{D}}}-
\frac{\left\lbrace 1-\frac{8(2\pi+\chi)\rho_0D^2}{3}\right\rbrace
}{\sqrt{1-\frac{4(2\pi+\chi)\rho_0D^2}{3}}}\right]}{2\left[\sqrt{1-\frac{4(2\pi+\chi)\rho_0D^2}{3}}-\sqrt{1-\frac{2M}{D}}\right]}.
\label{eq50}
\end{equation}

For $\omega(D)$ to be real it requires $\frac{2M}{D} < 1$ as well
as $\frac{4(2\pi+\chi)\rho_0D^2}{3} < 1$. Moreover, if one expands
the square-root terms in the numerator and the denominator of the
expressions of Eq. (\ref{eq50}) under the conditions
$\frac{M}{D}\ll 1$ and $\frac{4(2\pi+\chi)\rho_0D^2}{3}\ll 1$ in a
binomial series and retains the terms up to the first order, then
one can get
\begin{equation}
\omega(D)\approx
\frac{3}{2[\frac{3M}{2(2\pi+\chi)\rho_0D^3}-1]}.\label{eq51}
\end{equation}

Now, if one examines the above expression for $\omega(D)$ then two
possibilities may emerge out: either $\omega(D)$ is positive if
$\frac{M}{D^3}>\frac{2(2\pi+\chi)\rho_0}{3}$, or $\omega(D)$ is
negative if $\frac{M}{D^3}<\frac{2(2\pi+\chi)\rho_0}{3}$.

\section{Conclusion}
In the present work we have proposed a unique stellar model under
the $f(R,\mathcal{T})$ gravity as originally conjectured by
Mazur-Mottola~\cite{Mazur2001,Mazur2004} in the framework of
general relativity. The stellar model which they termed as gravastar,
may be considered to be a viable alternative to the black hole. To
fulfill the criteria of a gravastar they described the spherically
symmetric stellar system by the three different regions: interior
core region, intermediate thin shell, and exterior spherical
region with specific EOS for each of the region. Under this type
of specification we have found out a set of exact and
singularity-free solution of the gravitationally collapsing system
which presents several interesting properties which are physically
viable within the framework of alternative gravity of the form
$f(R,\mathcal{T})$.

In studying the above mentioned structural form of a gravastar we have noted down several salient aspects of the solution set as can be described below:\\
\indent (1) {\bf Pressure-density profile}: The pressure and
density relationship ($p=\rho$) of the ultrarelativistic fluid in
the shell is shown with respect to the radial coordinate $r$ in
Fig. 1 which maintains a constant variation throughout the shell.

(2) {\bf Proper length}: The proper length $\ell$ of the shell as
plotted with respect to the thickness of the shell $\epsilon$ (in
Fig. 2) shows a gradual increasing profile.

(3) {\bf Energy content}: The energy of the shell is directly
proportional to the thickness of the shell $\epsilon$ (in Fig. 3).

(4) {\bf Entropy}: The entropy $S$ within the shell has been
plotted with respect to the thickness of the shell $\epsilon$ (in
Fig. 4). This plot shows a physically valid feature that the
entropy is gradually increasing with respect to the thickness of
the shell $\epsilon$ and thus suggesting a maximum value on the
surface of the gravastar.

(5) {\bf Equation of state}: For $\omega(D)$ to be real it
requires $\frac{2M}{D} < 1$ as well as
$\frac{4(2\pi+\chi)\rho_0D^2}{3} < 1$. Moreover, under the
conditions $\frac{M}{D}\ll 1$ and
$\frac{4(2\pi+\chi)\rho_0D^2}{3}\ll 1$ upon expansion of the
expressions for $\omega(D)$ in a binomial series and retaining the
terms up to the first order two possibilities have been emerged
out: either $\omega(D)$ is positive if
$\frac{M}{D^3}>\frac{2(2\pi+\chi)\rho_0}{3}$, or $\omega(D)$ is
negative if $\frac{M}{D^3}<\frac{2(2\pi+\chi)\rho_0}{3}$.

Besides these important general features we have an overall
observation regarding the model in $f(R,\mathcal{T})$ gravity
which is as follows: unlike Einstein's general relativity there is
an extra term involving $\chi$ in the present model which has a
definite role and makes the fundamental differences between the
expressions in both the theories, as such vanishing of this
coupling constant $\chi$ provides a limiting case for getting back
the results of general relativity (e.g. note the
Ref.~\cite{Rahaman2015}). This aspect can be verified through a
comparative case study between the present work and that of Ghosh
et al.~\cite{Ghosh2017} under 4-dimensional background. In this
sense $f(R,\mathcal{T})$ gravity generates more generalized
solutions for gravastar than general relativity.

One final comment: as a possible astrophysical implication of our
results and tests to detect gravastars under $f(R,\mathcal{T})$
gravity one may study their gravitational lensing effects as
suggested by several authors, solely for
gravastars~\cite{Kubo2016} as well as for $f(R,\mathcal{T})$
gravity~\cite{Ahmed2016}. According to the methodology of Kubo and
Sakai one may adopt a spherical thin-shell model of a gravastar
developed by Visser and Wiltshire~\cite{Visser2004}, which
connects interior de Sitter geometry and exterior Schwarzschild
geometry. Now, assuming that its surface is optically transparent
they calculate the image of a companion which rotates around the
gravastar and find that some characteristic images appear,
depending on whether the gravastar possess unstable circular
orbits of photons (Model 1) or not (Model 2). For Model 2, Kubo
and Sakai calculate the total luminosity change, which is called
microlensing effects,where the maximal luminosity could be
considerably larger than the black hole with the same mass. In
future, if one study the similar effects under $f(R,\mathcal{T})$
gravity, then one can comparethe effects of modified gravity on
the above mentioned tests with that of the results based on
general
theory of relativity.\\

\section*{Acknowledgments}
F. R. and S. R. are thankful for the support from the
Inter-University Centre for Astronomy and Astrophysics (IUCAA),
Pune, India, and for providing the Visiting Associateship under
which a part of this work was carried out. S. R. also thanks the
authorities of the Institute of Mathematical Sciences (IMSc),
Chennai, India for providing the working facilities and
hospitality under the Associateship scheme. F. R. is thankful to
DST-SERB (EMR/2016/000193), Govt. of India for providing financial
support. Special thanks are due to Debabrata Deb for valuable
suggestions which became fruitful in preparation of the
manuscript. We are very grateful to the anonymous referee for
several useful suggestions which
 have enabled us to revise the manuscript substantially.


\begin{thebibliography}{99}

\bibitem{Mazur2001} P. Mazur and E. Mottola, Report number: LA-UR-01-5067.

\bibitem{Mazur2004} P. Mazur and E. Mottola, Proc. Natl. Acad. Sci. USA {\bf 101}, 9545 (2004).

\bibitem{Visser2004}  M. Visser, and D. L. Wiltshire, Classical Quantum Gravity {\bf 21}, 1135  (2004).

\bibitem{Cattoen2005} C. Cattoen, T. Faber, and M. Visser, Classical Quantam Gravity {\bf 22}, 4189 (2005).

\bibitem{Carter2005} B. M. N. Carter, Classical Quantam Gravity {\bf 22}, 4551 (2005).

\bibitem{Bilic2006}  N. Bili\'{c} , G. B. Tupper, and R. D. Viollier, J. Cosmol. Astropart. Phys. 02 (2006) 013.

\bibitem{Lobo2006} F. S. N. Lobo,  Classical Quantum Gravity {\bf 23}, 1525 (2006).

\bibitem{DeBenedictis2006} A. DeBenedictis, D. Horvat, S. Iliji\'{c}, S. Kloster, and K. S. Viswanathan, Classical Quantum Gravity {\bf 23}, 2303 (2006).

\bibitem{Lobo2007}  F. S. N. Lobo, and A. V. B. Arellano, Classical Quantum Gravity {\bf 24}, 1069 (2007).

\bibitem{Horvat2007}  D. Horvat, and S. Iliji\'{c}, Classical Quantum Gravity {\bf 24}, 5637 (2007).

\bibitem{Cecilia2007} C. B. M. H. Chirenti and L. Rezzolla, Classical Quantum Gravity {\bf 24}, 4191 (2007)

\bibitem{Rocha2008} P. Rocha, R. Chan, M. F A. da Silva, and A. Wang, J. Cosmol. Astropart. Phys. 11 (2008) 010.

\bibitem{Horvat2008}  D. Horvat, S. Iliji\'{c}, and A. Marunovic, Classical Quantum Gravity {\bf 26}, 025003 (2009).

\bibitem{Nandi2009} K. K. Nandi, Y. Z. Zhang, R. G. Cai, and A. Panchenko, Phys. Rev. D {\bf 79}, 024011 (2009).

\bibitem{Turimov2009} B. V. Turimov, B. J. Ahmedov, A. A. Abdujabbarov, Mod. Phys. Lett. A {\bf 24}, 733 (2009).

\bibitem{Usmani2011} A. A. Usmani, F. Rahaman, S. Ray, K. K. Nandi, P. K. F. Kuhfittig, Sk. A. Rakib, and Z. Hasan, Phys. Lett. B {\bf 701}, 388 (2011).

\bibitem{Lobo2013}  F. S. N. Lobo and R. Garattini, J. High Energy Phys. 12 (2013) 065.

\bibitem{Bhar2014} P. Bhar, Astrophys. Space Sci., {\bf 354}, 2109 (2014).

\bibitem{Rahaman2015} F. Rahaman, S. Chakraborty, S. Ray, A. A. Usmani, and S. Islam, Int. J. Theor. Phys. {\bf 54}, 50 (2015).

\bibitem{Ri1998} A. G. Riess et al., Astron. J. {\bf 116}, 1009 (1998)

\bibitem{Perl1999} S. Perlmutter et al., Astrophys. J. {\bf 517}, 565 (1999)

\bibitem{Bern2000} P. de Bernardis et al., Nature {\bf 404}, 955 (2000)

\bibitem{Hanany2000} S. Hanany et al., Astrophys. J. {\bf 545}, L5 (2000)

\bibitem{Peebles2003} P. J. E. Peebles and B. Ratra, Rev. Mod. Phys. {\bf 75}, 559 (2003)

\bibitem{Paddy2003} T. Padmanabhan, Phys. Rep. {\bf 380}, 235 (2003)

\bibitem{clifton2012} T. Clifton, P. G. Ferreira, A. Padilla, and C. Skordis, Phys. Rep. {\bf 513}, 1 (2012)

\bibitem{harko2011} T. Harko, F. S. N. Lobo, S. Nojiri, and S. D. Odintsov, Phys. Rev. D {\bf 84}, 024020 (2011).

\bibitem{Das2015} A. Das, F. Rahaman, B. K. Guha, and S. Ray, Astrophys. Space Sci. {\bf 358}, 36 (2015).

\bibitem{Das2016} A. Das, F. Rahaman, B. K. Guha, and S. Ray, Eur. Phys. J. C {\bf 76}, 654 (2016).

\bibitem{moraes2014b} P. H. R. S. Moraes, Astrophys. Space Sci. {\bf 352}, 273 (2014).

\bibitem{moraes2015a} P. H. R. S. Moraes, Eur. Phys. J. C {\bf 75}, 168 (2015).

\bibitem{moraes2015b} P. H. R. S. Moraes, Int. J. Theor. Phys. {\bf 55}, 1307 (2016).

\bibitem{singh2014} C. P. Singh and P. Kumar, Eur. Phys. J. C {\bf 74}, 3070 (2014).

\bibitem{rudra2015} P. Rudra, Eur. Phys. J. Plus {\bf 130}, 66 (2015).

\bibitem{baffou2015} E. H. Baffou, A. V. Kpadonou, M. E. Rodrigues, M. J. S. Houndjo, and J. Tossa, Astrophys. Space Sci. {\bf 356}, 173 (2015).

\bibitem{shabani2013} H. Shabani and M. Farhoudi, Phys. Rev. D {\bf 88},  044048 (2013).

\bibitem{shabani2014} H. Shabani and M. Farhoudi, Phys. Rev. D {\bf 90}, 044031 (2014).

\bibitem{sharif2014b} M. Sharif and M. Zubair, Astrophys. Space Sci. {\bf 349}, 457 (2014).

\bibitem{reddy2013b} D. R. K. Reddy and R. S. Kumar, Astrophys. Space Sci. {\bf 344}, 253 (2013).

\bibitem{kumar2015} P. Kumar and C. P. Singh, Astrophys. Space Sci. {\bf 357}, 120 (2015).

\bibitem{shamir2015} M. F. Shamir, Eur. Phys. J. C {\bf 75}, 354 (2015).

\bibitem{Fayaz2016} V. Fayaz, H. Hossienkhani, Z. Zarei, and N. Azimi, Eur. Phys. J. Plus {\bf 131}, 22 (2016).

\bibitem{sharif2014} M. Sharif and Z. Yousaf, Astrophys. Space Sci. {\bf 354}, 2113 (2014).

\bibitem{noureen2015} I. Noureen and M. Zubair, Astrophys. Space Sci. {\bf 356}, 103 (2015).

\bibitem{noureen2015b} I. Noureen and M. Zubair, Eur. Phys. J. C {\bf 75}, 62 (2015).

\bibitem{noureen2015c} I. Noureen, M. Zubair, A. A. Bhatti, and G. Abbas, Eur. Phys. J. C {\bf 75}, 323 (2015).

\bibitem{zubair2015a} M. Zubair and I. Noureen, Eur. Phys. J. C {\bf 75}, 265 (2015).

\bibitem{zubair2015b} M. Zubair, G. Abbas, and I. Noureen, Astrophys. Space Sci. {\bf 361}, 8 (2016).

\bibitem{Ahmed2016} A. Alhamzawi and R. Alhamzawi, Int. J. Mod. Phys. D {\bf 25}, 1650020 (2016).

\bibitem{Moraes2016} P. H. R. S. Moraes, J. D. V. Arba{\~n}il, and M. Malheiro, J. Cosmol. Astropart. Phys. 06 (2016) 005.

\bibitem{Yousaf2016a} Z. Yousaf, K. Bamba, and M. Z. H. Bhatti, Phys. Rev. D {\bf 93}, 064059 (2016).

\bibitem{Yousaf2016b} Z. Yousaf, K. Bamba, and M. Z. H. Bhatti, Phys. Rev. D {\bf 93}, 124048 (2016).

\bibitem{barrientos2014} O. J. Barrientos and G. F. Rubilar, Phys. Rev. D {\bf 90}, 028501 (2014).

\bibitem{singh2015} V. Singh and C. P. Singh, Astrophys. Space Sci. {\bf 356}, 153 (2015).

\bibitem{zeldovich1972} Y. B. Zel'dovich, Mon. Not. R. Astron. Soc. {\bf 160}, 1 (1972).

\bibitem{carr1975} B. J. Carr, Astrophys. J. {\bf 201}, 1 (1975).

\bibitem{Madsen1992} M. S. Madsen, J. P. Mimoso, J. A. Butcher, and G. F. R. Ellis, Phys. Rev. D {\bf 46}, 1399 (1992).

\bibitem{wesson1978} P. S. Wesson, J. Math. Phys (N.Y.) {\bf 19}, 2283 (1978).

\bibitem{braje2002} T. M. Braje and R. W. Romani, Astrophys. J. {\bf 580}, 1043 (2002).

\bibitem{linares2004} L. P. Linares, M. Malheiro, and S. Ray, Int. J. Mod. Phys. D {\bf 13}, 1355 (2004).

\bibitem{Israel1966} W. Israel, Nuovo Cimemto {\bf 44}, 1 (1966); {\bf 48}, 463(E) (1967).

\bibitem{Darmois1927} G. Darmois, “M{\'e}morial des sciences math{\'e}matiques XXV”,
Fasticule XXV, (Gauthier-Villars, Paris, France, 1927), chap. V.

\bibitem{lanczos1924} K. Lanczos, Ann. Phys. (Berlin) {\bf 379}, 518 (1924).

\bibitem{sen1924} N. Sen, Ann. Phys. (Berlin) {\bf 378}, 365 (1924).

\bibitem{perry1992} G. P. Perry and R. B. Mann, Gen. Relativ. Gravit. {\bf 24}, 305 (1992).

\bibitem{lake1996} P. Musgrave and K. Lake, Classical Quantum Gravity {\bf 13}, 1885 (1996).

\bibitem{rahaman2006} F. Rahaman, M. Kalam, and S. Chakraborty, Gen. Relativ. Gravit. {\bf 38}, 1687 (2006).

\bibitem{rahaman2009} F. Rahaman, M. Kalam, and K. A. Rahman, Acta Phys. Pol. B {\bf 40}, 1575 (2009).

\bibitem{usmani2010} A. A. Usmani, Z. Hasan, F. Rahaman, Sk. A. Rakib, S. Ray, and P. K. F. Kuhfittig, Gen. Relativ. Gravit. {\bf 42}, 2901 (2010).

\bibitem{rahaman2010} F. Rahaman, K.A. Rahman, Sk. A. Rakib, and P. K. F. Kuhfittig, Int. J. Theor. Phys. {\bf 49}, 2364 (2010).

\bibitem{dias2010} G. A. S. Dias and J. P. S. Lemos, Phys. Rev. D {\bf 82}, 084023 (2010).

\bibitem{rahaman2011} F. Rahaman, P. K. F. Kuhfittig, M. Kalam, A. A. Usmani, and S. Ray, Classical Quantum Gravity {\bf 28}, 155021 (2011).

\bibitem{Ghosh2017} S. Ghosh, F. Rahaman, B. K. Guha, and S. Ray, Phys. Lett. B {\bf 767}, 380 (2017).

\bibitem{Kubo2016} T. Kubo and N. Sakai, Phys. Rev. D {\bf 93}, 084051 (2016)

\end{thebibliography}
\end{document}